\documentclass[NPG]{copernicus}
\usepackage{graphicx}
\usepackage{color}

\definecolor{blue}{rgb}{0.00,0.00,1.00}
\definecolor{green}{rgb}{0.00,1.00,0.00}
\definecolor{red}{rgb}{1.00,0.00,0.00}
\definecolor{purple}{rgb}{0.63,0.13,0.94}
\definecolor{yellow}{rgb}{0.5,.50,.0}
\definecolor{orange}{rgb}{1.00,.50,0.00}

\begin{document}
\title{Thermal Energy Generation in the Earth}

\author[1]{Frederick J. Mayer}
\author[2]{John R. Reitz}

\affil[1]{Mayer Applied Research Inc.,
1417 Dicken Drive,
Ann Arbor, MI 48103}
\affil[2]{2865 S. Main Street (138),
Ann Arbor, MI 48103}


\runningtitle{Thermal Energy}
\runningauthor{Mayer and Reitz}
\correspondence{F. J. Mayer\\ fmayer@sysmatrix.net}

\received{}
\pubdiscuss{} 
\revised{}
\accepted{}
\published{}


\firstpage{1}

\maketitle  

\begin{abstract}
We show that a recently introduced class of electromagnetic composite particles can explain some discrepancies in observations involving heat and helium released from the Earth. Energy release during the \em{formation}\rm\,  of the composites and subsequent nuclear reactions involving the composites are described that can quantitatively account for the discrepancies and are expected to have implications in other areas of geophysics -  for example, a new picture of heat production and volcanism in the Earth is presented.
 \end{abstract}


\section{Introduction}
The early history of the heat from the Earth has been clearly reviewed in \citet{Cars}. After the earliest efforts in heat-flow modeling gave cooling times far too short compared to geological times, new developments in nuclear science gave rise to a better understanding of the effects of billion year time-scales of the radioactive isotopes of Uranium and Thorium.  This led, in-turn, to the now-familiar association between energy release by such isotopes and the heat from the Earth, [\cite{Dickin} pg 9]. We will refer to this association as the \textit{standard earth energy paradigm} or (SEEP).
 Although this picture is now-widely accepted, it has been shown to have some long-standing discrepancies with geophysical observations; see in particular,  
\cite{O&O1}, \cite{O&O2}, and \cite{v&K}. That there is heat being released from the Earth is not in question. Rather, the assumed heating due primarily to the long-lived alpha-particle decay of Uranium and Thorium has what van Kekan, et al., have called a ``robust'' discrepancy. Namely, the amount of heat measured when compared to the helium observed is too large by a factor of about twenty.
\par
 Other geophysical observations also bring into question the validity of the natural (U and Th) radioactive elements as the source of the Earth's heat flow. In particular the lack of a strong correlation of the geological depths and the heat flux measurements of \citet{marescal}, not to mention the difficulty in quantifying the absolute or relative amounts of the natural radioactive elements in the Earth,  [\citet{vans}], adding to the suspicion that there may be something wrong with the SEEP.
\par In addition to the heat to helium imbalance, geophysical measurements of the helium released also appear to be peculiar. The vented helium usually has a small amount of $^{3}\mathrm {He}$ present in the dominantly $^{4}\mathrm {He}$ amounting to about one part in $10^{5}$ except in the vicinity of volcanically active sites where the ratio may be larger by two or more orders of magnitude. However, in contrast to the $^{4}\mathrm {He}$, there is no source of $^{3}\mathrm{He}$ from radioactivity, hence, geophysicists have assumed it to be primordial, continuing to be vented along with the $^{4}\mathrm{He}$. But, the $^{3}\mathrm {He}/^{4}\mathrm {He}$ ratio is almost the same around the Earth when sampled at the mid-ocean ridges. This observation argues for a common, or perhaps a connected, source for the helium isotopes as has been suggested by {\citet{Hern}.\,We point out that the statistics in Herdon's Table 1 shows that the ratio  $^{3}\mathrm{He}/^{4}\mathrm {He}$\,on the mid-ocean ridges (far from volcanic activity) varies by no more than about a factor of two at about $10^{-5}$ in these locations}.  More will be said about this data in later sections.
\par 
We will not try to address all of the issues with the SEEP but rather to focus on quantitative questions relating to helium releases, noting that \cite{O&O2} started along this path many years ago. 
\par
The heat to $^{4}\mathrm {He}$ imbalance problem might be removed if there was another source generating both heat and helium. In this paper, we propose and describe that there is such a source. As we show, the recently introduced Compton composite particles  [\citet{M&Ra}, hereafter (M\&Ra)]  provide an alternative mechanism for heat production, give a somewhat different picture of how volcanic activity is triggered in the Earth, and shows how $^{3}\mbox{He}$ and $^{4}\mbox{He}$ are both generated in the process.  The new composites, called \em{tresinos}\rm, are made up of a hydrogen nucleus (proton,  deuteron, or  triton) and two electrons bound together by electromagnetic forces; they are roughly ten times smaller than hydrogen atoms.  Their energies of \em{formation}\rm, \,i.e., their binding energies of approximately 3.7 kilo-electron volts, is the dominant source of the new heat as we show below. 
\par 
Our picture of volcanic activity is not that of heat conducting (or percolating) up from deep in the Earth, and then being impeded by a region of low thermal conductivity until it finally breaks through.  Rather, it is generated at 50 to 200 km depths by hydrogen nuclei from intercalated acidic water, combining with electron pairs from oxide ions (e.g., carbonate, sulfate, and/or alumino-silicate minerals) under elevated temperature and pressure.  It is not surprising that most volcanic regions are located on the edges of the continents close to the oceans or other sources of water (note the map at the USGS website: vulcan.wr.usgs.gov). 
\par 
But what is the origin of Earth's outgassing helium, especially $^{3}\mbox{He}$?  Earth's water contains a small amount of deuterium so deuteron \em{tresinos}\rm \, will be produced along with proton \em{tresinos}\rm. Common and well-understood fusion  reactions (e.g., \em{d-d, d-t}\rm) which result in helium production [see e.g. \citet{R&R}, pg 338] can not be considered because the temperatures in the Earth are much too low for these reactions to take place. However, the recently proposed Compton composite particles (M\&Ra), allows a unique and remarkable group of nuclear reactions to be generated. In fact, a chain of reactions with these particles in the low temperatures of the Earth become more than possible --  they seem to be necessary to explain many otherwise quite paradoxical geophysical measurements.
\par
In this paper, we hope to explain a number ``strange" geophysical observations (difficult to explain within the SEEP), that result from the Composite particle formations and later interactions, while otherwise going unnoticed.
\par
The outline of this paper is as follows. In Section 2, we present an \textit{Introduction to Tresino Physics}. In Section 3, how \textit{Tresinos in the Earth} can be formed. In Section 4, we introduce \textit{The Deuteron Tresino Nuclear Chain} arriving at the system of rate equations representing the primary nuclear reactions. In Section 5, we present our \textit{Numerical Results} derived using the formulated rate equations assuming a very large reaction zone. In Section 6, we discuss \textit{Tresino Formation and the Heat to Helium Balance} paradox. Subsection 6.1 discusses recent geoneutrino experiments. We finish, in Section 7, with a \textit{Discussion and Conclusions} comparing our picture of tresino-induced Earth heating and some other possible implications of tresino generation in geophysics. 
\par 
Appendix A1 presents the derivation of the \textit{Deuteron Tresino Reaction Rates}. Appendix A2 presents the derivation of the \textit{Infinite-Medium Reaction Rate Equations}. Appendix B presents \textit{A Simple Earth Heat Flow Calculation} originating at modest depths. 
\section{Introduction to Tresino Physics}
The electromagnetic composites, we call \em{tresinos}\rm,
have been described in our recent paper (M\&Ra). The tresino is a unique and strongly bound composite particle.  It is much more strongly bound than the weakly bound negative hydrogen ion composed of the same three particles. Because it is so strongly bound, the tresino is considerably smaller than an atom and it does not interact chemically, in the usual way, with atoms or molecules. It may be helpful for readers unfamilair with nuclear reaction physics to look over our recent paper before trying to understand the application of tresinos in geophysics presented in this paper.
\par
Tresinos are Compton-scale composites composed of two electrons and a hydrogen nucleus bound together by electromagnetic forces.  That is to say, the dimensions are roughly at the electron Compton wavelength (\mbox{$\lambda_{c} \approx 3.8\times 10^{-11}$} cm); because this is the natural dimensional scale of these particles, all dimensions are presented in Compton units.
\mbox{Figure 1} illustrates the tresino configuration in a ``classical'' picture; a more realistic quantum mechanical description is presented in our paper (M\&Ra).
\begin{figure}[h]
\begin{center}
\resizebox{5cm}{1.666cm}{
\includegraphics{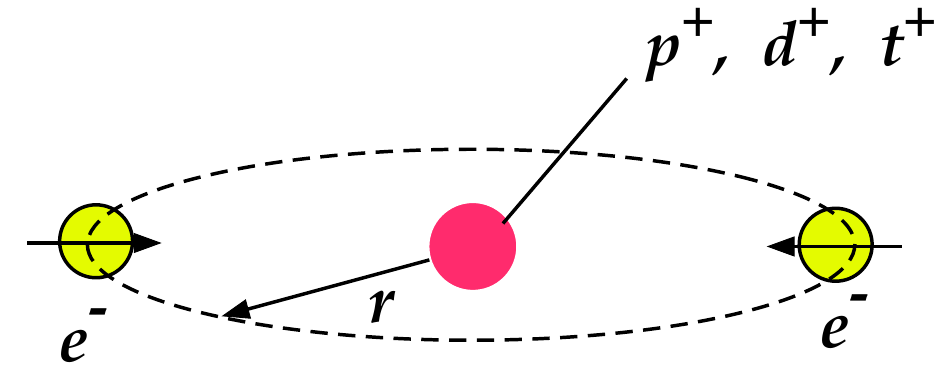}}
\caption{The ``classical'' tresino electromagnetic configuration. Note that tresinos are quantum objects so the distance, on average, between the electrons (in {\color{yellow}{yellow}}) is about $15$\,Comptons ($\lambda_{c}$'s.)}
\end{center}
\end{figure}
It is important to note that the tresino has a net negative charge and is quite small (roughly a factor of ten smaller than the hydrogen atom). It should be clear that tresinos will behave like heavy, negatively-charged ``ions" having approximately the mass of the hydrogen nucleus; tresinos are electrostatically attracted to positive charges. Because they are three-body electromagnetic entities, assembled from a hydrogen nucleus and two electrons, tresinos are not either easily, or usually, formed. Upon \em{formation}\rm, they release their binding energy of $E_{b}=3.7$ keV and are stable unless the binding energy is, in some way, re-supplied to make them disassemble; this is a substantial amount of energy on the scale of usual chemical reactions (less than a few eV). The tresino \em{formation}\rm \,\,and re-ionization processes are similar to the ionization and recombination of a hydrogen atom, except that they are, in both directions, much more energetic.  
\par
In preparation for what we discuss below, we present an overview of tresino physics because these particles will represent a new concept to most Earth scientists and physicists   generally. We discuss some specific characteristics, that we hope will clarify questions that could arise while we will try to avoid too much repetition in the derivations and model calculations presented in later sections.
\par
First of all, tresinos are not easily produced or detected. The conditions under which they will form, and release their binding energies, are unique and complex, i.e., compared to usual chemical reactions due to their three-body composition. Upon \em{formation}\rm, \,the released binding energy goes into kinetic energy of recoiling particles. In a tresino formation collision, another ``heavy" partner nucleus must be involved in order to conserve momentum in the formation energy release.
The $3.7$ keV binding energy is shared between the recoiling tresino and the partner  nucleus. This means that a tresino and its close-by partner share the binding energy between them and hence both are given a substantial kinetic energy ``kick" on the scale of usual chemical reactions, typically only a fraction of an eV.
\par
Because tresinos are negatively charged and stable they will immediately be attracted to a close-by positively charged ion or nucleus. Because the kinetic energies of the tresino and its recoil partner are at the keV level (i.e., high on the scale of chemical energies but low on the scale of nuclear energies) they lose this energy over a short distance at solid densities, depositing the recoil energy but otherwise being effectively hidden from easy view or, for that matter, easily detected experimentally. This characateristic makes direct detection of tresinos in experiments difficult; tresino generation reveals its presence almost exclusively by the heating that accompanies its formation. 
\par 
Next, we turn to the question of how tresinos might be formed from common constituents (atoms and ions) in the Earth.
\section{Tresinos in the Earth}
There are many materials, such as alumino-silicate clays with desolved hydrogen ions, that may be conducive to tresino formation. A heated clay (possibly molten), montmorillonite for example, contains many oxygen ions. Such clays have high-densities of localized electron pairs in the form of oxide ions $\mathrm{\mbox{O}}^{2-}$. Of course, other phyllosilicates and salts are common in magmas as well. With hydrogen ions, dissolved from sea water, in such a ``soup'' of oxides, tresinos may \textit{self-assemble} in reactions or collisions. A ``soup" of chemicals containing both hydrogen ions, most probably as $\mathrm{H}_3 \mathrm{O}^{+}$ ({\em{hydronium}\rm) \,alone or hydrated, and numerous ions having pairs of available electrons, clearly this composition will favor tresino generation.
\par
We first note that in this environment, since the water comes from the sea, there will be both types of hydrogen ions present, protons (from ``light water", H$_{2}$O) and deuterons (from ``heavy water", D$_{2}$O and HDO).  We can estimate the tresino formation rate (either proton or deuteron types) by considering the collision of the proton or deuteron with, say, an $\mathrm{O}^{2-}$ ion. Moreover, as there can be many different ion species with the required pairs of unbound electrons, we proceed by considering a ``generic" donor delivering the electron pairs to form the tresinos. For purposes of the following discussion, we  consider the oxygen ion as the donor, keeping in mind that other such ions can play the same role. (Note: from here on we simplify the notation by denoting a tresino by an \mbox{asterisk {*}}, i.e., $p^*$\, $d^*$, $t^*$. \mbox{Figure 2} illustrates a proton collision with an $\mathrm{O}^{2-}$ ion generating a proton tresino. Obviously, a similar collision will generate deuteron tresinos from deuterons. 
\begin{figure}[h]
\centering
\resizebox{8cm}{3.2cm}{
\includegraphics{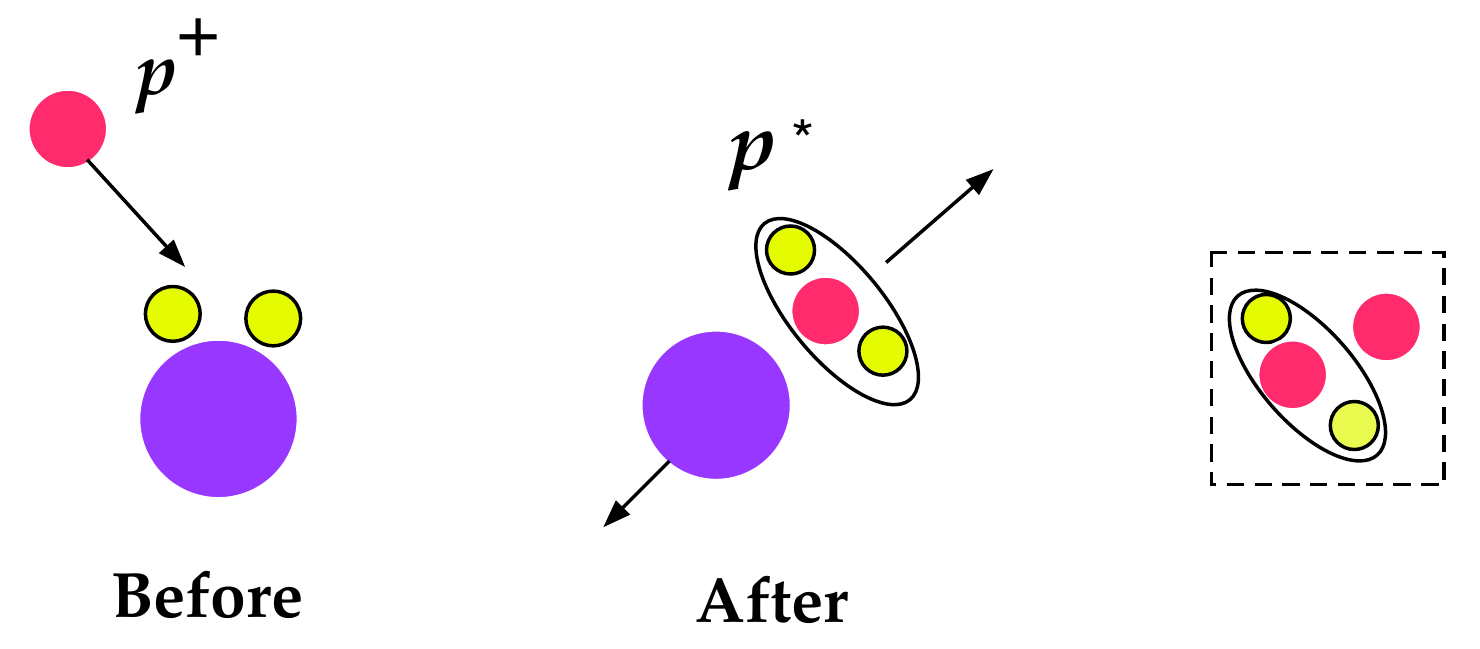}}
\caption{A illustration showing a formation collision of a proton (in {\color{red}{red}}) and an $\mathrm{{O}}^{2-}$ ion (in {\color{purple}{purple}}), the latter having two unbound electrons (in {\color{yellow}{yellow}}).   A PTM is illustrated in the box (see text for explanation).}  
\end{figure}
Furthermore, note that both collision partners may simply be imbedded as ions of a clay or salt structure with enough vibration to bring them into collisional contact. 
\footnote {Note that in a very different context [\cite{M&Rb}], we have considered tresino formation process in which a three-body resonant-scattering (an electron sandwiched between two protons) forms first, and is then quickly followed by another electron collision. Such a situation might also occur in the Earth under certain chemical/physical situations (a comlex three-body interaction).}
\par 
Now, let us estimate the tresino generation rate in such collisions. A simple estimate of the  generation rate of proton tresinos $p^{*}$ can be written, 
\begin{equation}
\hspace{2cm} dn_{p^{*}}/dt \approx \eta\, \sigma v\,n_{p}\,n_{\mathrm{O}^{2-} }
\end{equation} 
where we take the collision cross-section to be $\sigma\approx\pi r_{ion}^{2}$, with $r_{ion}$ the effective oxygen ion radius,  $v$ is the proton velocity, and $\eta$ is the fraction of such collisions that results in the capture of a pair of oxygen-ion electrons. We expect the value of $\eta$ to be quite small, due primarily to size differences and statistical considerations. We will consider $\eta$ to be a parameter; however, looking ahead to later numerical solutions, we will choose a value which gives the correct order of magnitude for the \mbox{``excess heat"} generated in so-called \mbox{``cold fusion"} experiments of  \cite{Notoya}. These experiments involved a solution containing both protons as well as other ions. As discussed in M\&Ra, and noted above, the formation of a tresino releases a large (on the scale of chemical reactions) amount of energy (3.7 keV), which deposits in the medium in which the formation process takes place. Also in M\&Ra, we had suggested that the excess heat measured by Notoya and others, was most likely an example of $p^{*}$\,formation energy release.
\par 
We do not mean to suggest that a molten clay or salt is required with dissolved hydrogen in order to produce tresinos.  The Earth's crust and upper mantle are mostly composed of silicate materials, and some silicates can absorb some wt\% of water and acid.  It is widely thought (by geophysicists) that there is a substantial amount of water stored in these upper regions of the Earth -- \citep{J&L} and \citep{R&G}.  Although some is undoubtedly absorbed as water of hydration or as hydroxyl units, it is also believed that some of the hydrogen enters silicate materials as $\mbox{H}^{+}$ or $\mathrm{H}_3\mathrm{O}^+$ hydronium ions, see  \citep{R&G}, \citep{H&B}, and \citep{liu}.  In fact, the latter investigators have identified potential ``docking sites" for the $\mbox{H}^{+}$ ions, all of which are located close to $\mbox{O}^{2-}$ ions.  Thus, solid silicate materials may be a source of tresinos under the elevated temperature and pressure conditions at modest depths in the Earth. 
\par As already mentioned, there may be many types of electron-pair \textit{donors}. So, in order to simplify the notation, from here on we denote all electron-pair donors, like the $\mathrm{{O}}^{2-}$ ion, with the subscript\,\,\bf{\em{\large{\color{yellow}{ee}}}}\rm.
\par 
After a proton tresino is generated, it carries most of the binding energy away as kinetic energy and slows down in a short distance (several microns) in the host material. After depleting the kinetic energy, the tresino ultimately captures an ambient proton and spins down and becomes a \em{Proton Tresino Molecule}\rm\, or PTM, for short. It is \em{like}\rm\, a common molecule except that it is much smaller and not bound together by the same forces. A schematic of a PTM is illlustrated in the box in \mbox{Figure 2}. Note that PTMs are neutral, quite small ($\approx $15 Comptons), and will not interact with ordinary atoms or ions; therefore they move easily through ordinary matter, escape into the atmosphere, and are ultimately lost from the Earth just like an ordinary hydrogen molecule is. 
\section{The Deuteron Tresino Nuclear Reaction Chain}
We concentrate on the so-called ``infinite-medium" situation -- meaning that all boundaries are sufficiently far away so no participating particles either enter or leave the reaction zone. This is not a significant limitation because all of the energetic reacting particles involved in the system have rather short mean-free paths at solid, or near-solid, densities.
\par
The generation of proton and deuteron tresinos and the release of their binding energy to the medium is just the beginning of thermal energy generation. In this ionic environment, having proton and deuteron tresinos as well as protons and deuterons, there will be many collisions that may initiate additional reactions, specifically, nuclear reactions.  The most important reactions are the \em{d-$d^{*}$}\rm\, reactions and the nuclear reaction chain they initiate. We will neglect the background \em{p-$p^{*}$\,}\rm and \em{p-$d^{*}$}\,\rm collisions (incoming and exiting particles are identical) and concentrate on the \mbox{\em{d-$d^{*}$}\rm}\,collisions and nuclear reaction chain resulting therefrom.  Specifically, \mbox{a $d^{*}$ may collide with an ordinary $d$.} These two particles are electrostatically accelerated into each other until a distance of closest approach is reached. At this point, the tresino shielding may be lost and the two deuterons then will electrostatically repel. 
\begin{figure}[h]
\centering
\resizebox{6.66cm}{5cm}{
\includegraphics{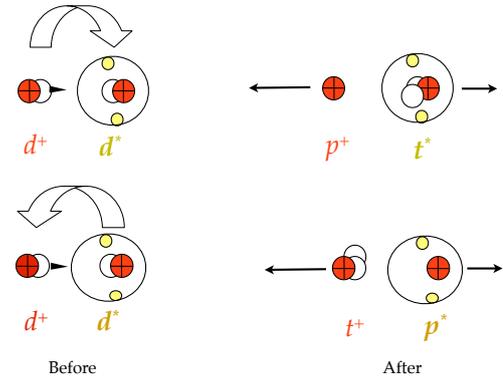}}
\caption{An illustration of the two possible \em{d-$d^{*}$}\rm\,neutron transfer reactions. Protons (in {\color{red}{red}}), neutrons colorless, and electrons (in {\color{yellow}{yellow}}).}
\end{figure}
However, in some cases, the distance of  closest approach is small enough to allow a neutron to be transferred between the two deuterons, thereby releasing 4 MeV of nuclear energy as kinetic energy of the recoiling particles. Note that a neutron may be transferred in either direction. Figure 3 illustrates these two collision-induced neutron transfer reactions; these reactions represent one branch of the well-studied $d$-$d$ fusion reaction, see \citet{R&R} pg 338, and are just the first reactions in a chain of deuteron tresino reactions.  In reaction notation, they may be written,
\begin{equation}
   \hspace{2cm}  d\,+ \, d^{*}\rightarrow p\,+\, t^{*}\,+ 4 \mbox{\,\,MeV} 
\end{equation}
\begin{equation}
  \hspace{2cm}  d\,+\, d^{*}\,\rightarrow p^{*}\,+\, t\, + 4 \mbox{\,\,MeV}.
\end{equation}
Note that these reactions have \em{self-assembled}\rm\, due to their electrostatic acceleration at large separations and the shielding provided by the tresino electrons at small separations, i.e., less than about a Compton.  However, the only nuclear reactions that occur are via neutron transfers sometimes described as\,``under the Coulomb barrier'',  [\cite{vol}].  We emphasize that this is a rather unique situation because the two nuclei do not collide to form a compound nucleus, i.e., a \em{fusion}\rm \, reaction. But they do come close enough together to allow a neutron transfer that releases the nuclear energy from just this one branch of the $d$-$d$ \em{fusion}\rm \,\,reaction. The result of this special class of nuclear reactions is that only charged particles are produced and will carry away the reaction energy as recoil kinetic energy of the exiting particles. 
\par In reactions (2) and (3), both the triton $t$ and the triton tresino $t^{*}$  must eventually $\beta$-decay releasing a neutrino and an electron as is well-known [\cite{blw} pg 219]. 
The ``\em{free}\rm"\, triton indicated in reaction (3) will be quickly neutralized by picking up an ambient electron as it slows down. On the other hand, the triton tresino $t^*$ in reaction (2) is protected from neutralization until it $\beta$-decays and becomes 
``exposed" as a $^{3}\mbox{He}$ nucleus whereupon the $t^{*}$ is broken-up by the energetic electron ejected in the $\beta$-decay process.
\begin{figure}[h]
\centering
\resizebox{8.54cm}{2.9cm}{
\includegraphics{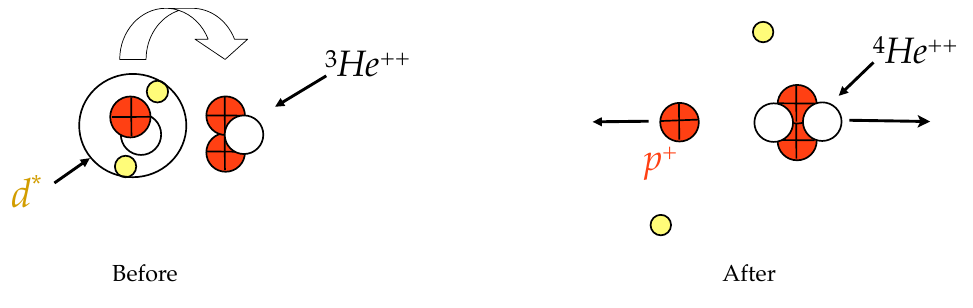}}
\caption{An illustration of the deuteron tresino $d^{*}$-$^{3}$He neutron transfer reaction. Protons (in {\color{red}{red}}), neutrons colorless, and electrons (in {\color{yellow}{yellow}}).}
\end{figure}
With the ``exposed" $^{3}\mbox{He}$ nucleus, a possibility then exists for another neutron transfer reaction with another deuteron tresino $d^{*}$. This reaction is both more energetic and of higher-probability (a larger reaction cross-section); in reaction notation it is,
\begin{equation}
\hspace{2cm}    d^{*}\, +\, ^{3}\mbox{He} \,\rightarrow \, p\, + \, ^{4}\mbox{He}\, + 18
    \mbox{\,\,MeV}.
    \end{equation} 
Again, this reaction is a well-studied neutron transfer reaction (see Rolfs and Rodney, p. 338).  
Figure 4 illustrates this reaction. Note that this energetic reaction removes the electrons that were previously bound in the incoming deuteron tresino $d^{*}$. Exiting this reaction, the proton and $^{4}$\mbox{He} nucleus ($\alpha$-particle), share 18 MeV as recoil kinetic energy. 
\par
Note from Figures 3 and 4, that there are now more energetic particles that are involved in the ``soup" of particles. In particular, there are now energetic tritons $t$'s,  triton tresinos $t^{*}$'s, protons, $p$'s, $^{3}\mbox{He}$ nuclei, and $^{4}\mbox{He}$ nuclei, having been given extra kinetic energy from the nuclear reactions driven by this chain of deuteron tresino reactions. 
\par So in summary, at the end of this chain of \em{self-assembled}\rm\, neutron 
transfer reactions, we find approximately 22 MeV of nuclear energy released for each\,$^{4}\mbox{He}$ (\mbox{$\alpha$-particle}) produced. This should  be compared to only about 5 MeV released for each radioactive $\alpha$-particle decay of U and Th nuclei. Moreover, the nuclear chain reactions just described, result in the energy being released as charged-particles that deposit their energy close to their point of origin.  In addition to the energy released in the nuclear reactions, there is a substantial amount of tresino formation energy released, in an even shorter mean-free path. 
\par
Because the triton $t$ and triton tresino $t^{*}$ both have 12.3 year half-lives, these decays take place over a substantial period of time; on the other hand, there is a lot of time available in Earth-related processes. Because of this, the tritium half-life plays a big role in the net energy release as we proceed to show. 
\par
The top portion of Figure 5 illustrates the proton tresino formation reaction; it  releases 3.7 keV in recoil kinetic energy. This is true for the deuteron tresino formations in the bottom portion of Figure 5 as well.  But further, the bottom portion illustrates the entire primary deuteron tresino reaction chain; it releases 22 MeV of recoil kinetic energy. Note that the exiting particles are indicated by outwardly directed arrows. The \mbox{$\alpha$-particle} and the proton, at the end of the chain, carry away most of the nuclear reaction energy. 
\par
It was important to formulate a quantitative numerical model for the net energy release and the numbers of nuclei of various types generated to facilitate comparisons with geophysical observations.  To do this, we derived the relevant reaction rates for the tresinos and other involved nuclei. The derivations are a straightforward modification of some conventional nuclear reaction rate calculations. Modifications were required due to the unique character of the deuteron tresino reactions, i.e., for the neutron transfer reactions.  
\begin{figure*}[h]
\begin{center}
\includegraphics[width=9cm]{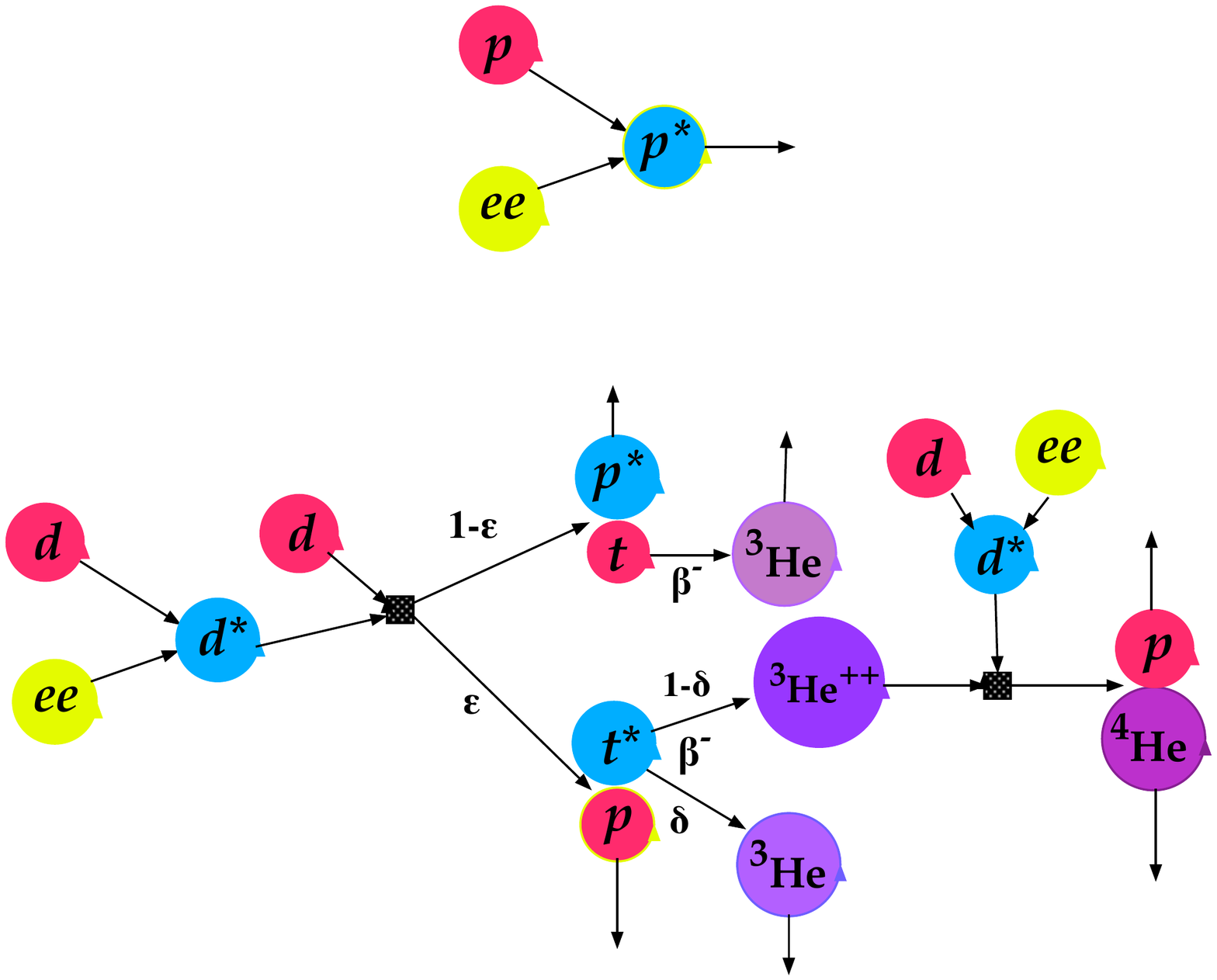} 
\end{center}
\caption{An illustration showing the proton tresino formation (top), releasing 3.7 keV, and the deuteron tresino reaction chain (bottom) releasing 22 MeV. Note that there are three deuterons ``consumed" for each $\alpha$-particle produced.}
\end{figure*}
\par
We present the relevant reaction rate derivations in Appendix A1 and the equations linking the various species in Appendix A2. It may be useful to examine the Appendices along with Figure 5 to clarify the interconnections of this reaction system.  Some constants of the formulation are also presented in the Appendices. 
\begin{center}
\bf{Table 1. The Infinite-Medium Model Rate Equations}\rm
\begin{displaymath}
\hspace{2cm}dn_{p^{*}}/dt = \eta\, \sigma v\,n_{p}\,n_{ee} 
\end{displaymath}
\begin{displaymath}
\hspace{2cm}dn_{d^{*}}/dt = \eta\, \sigma v\,n_{d}\,n_{ee}
\end{displaymath}
\begin{displaymath}
\hspace{2cm}dn_{{d}}/dt\, =\, -\,\,r_{d\,ee}\, - \, r_{d\,d^{*}}
\end{displaymath}
\begin{displaymath}
\hspace{2cm}dn_{\mathrm{ee}}/dt\, =\,-\,r_{d\,ee}\,-r_{p\,ee}
\end{displaymath}
\begin{displaymath}
\hspace{2cm}dn_{d^{*}}/dt\, =\, r_{d\,ee}\,-\, r_{d\,d^{*}}\,-\, r_{3\,d^{*}}
\end{displaymath}
\begin{displaymath}
\hspace{2cm}dn_{t}/dt\,= (1-\epsilon)\, r_{d\,d^{*}}\, - \,n_{t}/\tau
\end{displaymath}
\begin{displaymath}
\hspace{2cm}dn_{t^{*}}/dt\,= \,\epsilon \,\,r_{d\,d^{*}}\, -  \, 
    n_{t^{*}}/\tau
\end{displaymath}
\begin{displaymath}
\hspace{2cm}dn_{3}/dt\, =\,n_{t}/\tau\, + \,n_{t^{*}}/\tau\,-\, \,r_{3\,d^{*}}
\end{displaymath}
\begin{displaymath}
\hspace{2cm}dn_{4}/dt\,= \,r_{3\,d^{*}}
\end{displaymath}
\end{center}
\par In Table 1, we have pulled together the infinite medium reaction rate equations. The two branches of the deuteron tresino reactions are indicated by the fractions $\epsilon$ and $(1-\epsilon)$, and the tritium decay constant $\tau$ can be found in the Appendices. The parameter $\eta$ has been discussed in Section 3. This completes the set of non-linear reaction rate equations that represent the interconnected reactions to be solved for a given rate of input (acidic) water and density of available of donor ions \bf{\em{\large{\color{yellow}{ee}}}}\rm.
\par 
In the equations of Table 1, the dependent variables are all functions of time: $n_{ee}$ is the (electron pair) donor ion density, $n_{p}$ is the proton density, $n_{d}$ is the deuteron density, $n_{p^{*}}$ is the proton tresino density, $n_{d^{*}}$ is the deuteron tresino density, $n_{t}$ is the triton density, $n_{t^{*}}$ is the triton tresino density, $n_{3}$ is the $^3$He density, and $n_{4}$ is the $^4$He density. 
\par
The parameters $\epsilon$ and $\tau$ must be set before integrating the infinite-medim model equations.  First, the $\beta$ decay constant $\tau=17.7\,\mathrm{years}\rm$. Next, referring to \mbox{Figure 5}, it is clear that to produce ``free'' tritons there would have to be an\,$\epsilon$ value significantly different from one. However, in contrast to $^{3}\mathrm{He}$ and $^{4}\mathrm{He}$, there does not appear to be experimental evidence for tritium production from the Earth when geophysicists have examined magma from volcanoes [see\cite{Goff}, \cite{Jones}]. Even with the relatively short half-life of 12.3 years this argues for a value of $\epsilon$ close to one.  Therefore, for now, we will choose $\epsilon =1$ in the numerical evaluations. Note that a value less than one would not only generate tritium, it would generate additional $^{3}\mathrm{He}$.  Further notice that $\epsilon\simeq 1$ means that in the neutron transfer reaction, the neutron is transferred into the $d^{*}$ converting it \mbox{to a $t^{*}$} (see Figure 3).
\section{Model Rate Equation Integration: A Typical Example}
\par The numerical solution to this set of coupled non-linear ordinary differential 
equations is straightforward. They are easily programmed for examining various input parameter choices in the infinite-medium model. 
\par As an example of the numerical integrations of the model equations, we consider the situation where water flows into, or is entrained in, some generic molten clay containing \em{donor}\rm\,\,ions at a rate $\nu$\,(yr$^{-1}$) having the following (normalized) form:
\begin{center} 
$ n_{p}(4\,\nu/\sqrt{\pi})(\nu\,t)^{2}\exp[-(\nu\,t)^{2}]$
\end{center}
 where $n_{p}$ is the proton density that would have been achieved if 
no tresinos were to have been formed, thereby reducing their numbers. A similar model is taken for the deuteron density with $n_{d}=n_{p}/6600$, the sea-water value. We denote the starting electron donor density as $n_{ee}(0)$ that is depleted by the tresino formation reactions \mbox{(see Table 1)}.
\begin{figure*}[h]
\includegraphics[width=13cm]{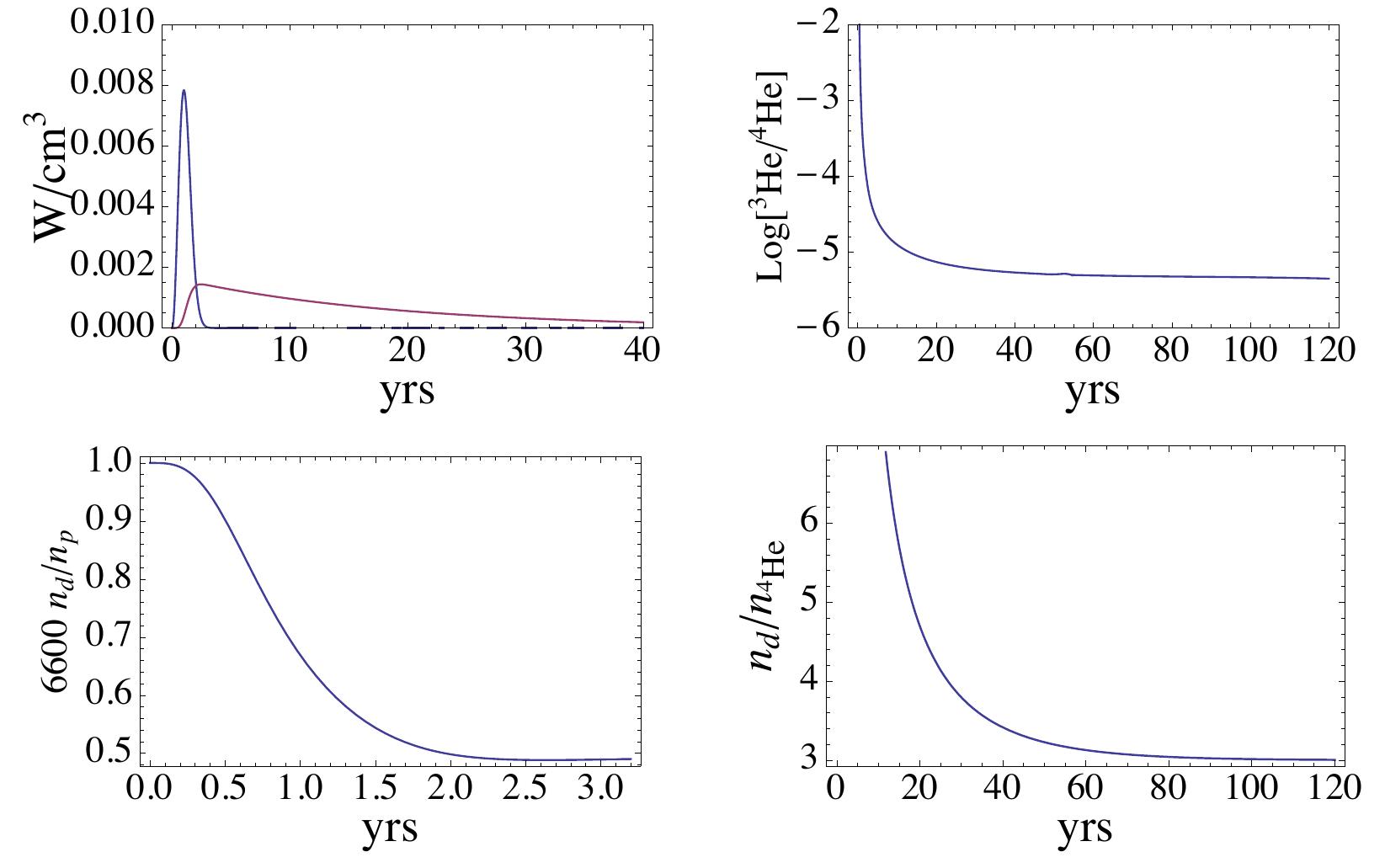}
\caption{Upper left panel: The proton tresino formation power (the fast 
``spike'') and the much slower nuclear reaction power (presented here multiplied by a 
factor of ten for graphical clarity). Upper right panel: The Log of ratio $^{3}\mathrm{He}/ ^{4}\mathrm{He}$.  Lower left panel:  The Log of the ratio $6600 \,\,n_{d}/n_{p}$ in the early time. Lower right panel: the ratio of initial deuteron density to the $^{4}$He density.}
\end{figure*}
\par  
Figure 6 displays a solution of the rate equations over a period of about ten tritium half-lives starting from the following parameters: 
\mbox{$n_{ee}(0)= 10^{21}\mathrm{cm}^{-3}$}, \mbox{$n_{p}(0)=5\times 
10^{20}\,\mathrm{cm}^{-3}$}, \mbox{$\nu=1$},  i.e., water is introduced over a period of one year,  \mbox{$\epsilon=1$}, \em{T}\rm\,=\,2000 $^{\mathrm{o}}$K, the ``tunneling" parameter \mbox{$P=2.3\times 10^{-5}$} (see Appendix A1), and $\eta=10^{-16}$ (see Section 3). In the upper left panel of \mbox{Figure 6}, we display the proton tresino formation power, it is the narrow ``spike'' and is roughly on the time-scale of the water inflow rate $\nu$ and also the very much longer time-scale and lower nuclear reaction power from the deuteron chain (the latter has been multiplied by a factor of ten for graphical clarity). In the upper right panel of Figure 6, we display the ratio of the helium isotope densities $^{3}\mathrm{He}/ ^{4}\mathrm{He}$ generated as a function of time. Here, there are two important points to notice: (i) the late-time ratio is about $10^{-5}$, and (ii) early in time the ratio is about a factor of a thousand larger. 
\par 
\citet{Clarke}, first observed a small amount of 
$^{3}\mathrm{He}$ in recovered helium gas from the Earth, a result that is
now well established. Furthermore, the measured value of the $^{3}\mathrm{He}$/$^{4}\mathrm{He}$ ratio has been found to be remarkably consistent (within a factor of two) at about $10^{-5}$ as already noted in the data presented by \cite{Hern}, away from geologically active sites, i.e., volcanoes. On the other hand, this ratio is found to be orders of magnitude larger near the active sites [\cite{Dickin}]. The relevant time-scale separating these different helium isotope ratios at the two locations, is as expected, the tritium decay time.
\par 
The lower left panel of Figure 6 shows the ratio of the unreacted deuterons to the 
 unreacted protons. Two deuterons are removed early in 
 tresino formation reactions, and a third deuteron (as a 
 $d^{*}$) is removed later in $^{4}\mathrm{He}$ generation. During 
 this same period, about 6600 protons are removed in tresino formation.
 Finally, the lower right panel of Figure 6 shows the ratio of starting deuteron density to the $^{4}$He density in late-time showing again that three deuterons are consumed for each $^{4}$He produced. 
\par In performing many model integrations, it was found that, for a 
given choice of the parameters, varying only the water inflow rate 
$\nu$ over orders of magnitude, the late-time helium isotope ratio 
varies less than about a factor of two. Thus, the late-time ratio of 
these isotopes is nearly always the same regardless of the water inflow rate. This result is in quantitative agreement with the data at the mid-ocean ridges mentioned in Section 1 (see Herdon's paper).
\par 
Another result from running the rate equation integrations is that there is 
a nearly linear relationship between the maximum tresino power 
generation and the water inflow rate - the higher the rate, the more power is 
generated. This is to be expected as the peak power is correlated with the 
proton tresino formation reactions. Of course, with a faster inflow rate, say $\nu \approx 20$, the thermal power level is increased to about \mbox{0.15 W/cm$^3$}. At this level, the power released from a cube roughly twenty meters on a side is a few terawatts in about 20 days. We note that this power level is sufficient to drive an enormous eruption like that of Krakatowa. 
\par It has been estimated that there are roughly 1500 active volcanoes around the Earth so, on average, less than 0.03 terawatts per volcano is require to account for the measured excess power from the Earth. At the peak power of 0.008 W/cm$^3$ of \mbox{Figure 6}, a cube of about 150 meters on a side will produce a 0.03 terawatt level, a power level easily produced in the tresino reaction chains. Much larger power levels are clearly possible. 
\par Although they have been very instructive, we know that the rate equation integrations for the reaction chains are not fully self-consistent -- this is because the temperature of the medium has been treated as a constant parameter. As reactions begin, the temperature will start to increase  and then the reaction rates will also increase as long as the donor and/or hydrogen nuclei have not been depleted. Then, at some higher temperature, the donor ions may be destroyed thus resulting in a self-limiting effect upon the power generation of the tresino reaction chains. 
\section{Tresino Formation and the Heat to Helium Balance}
In addition to the energy from the nuclear reaction chain, as we have shown, there is substantial energy of formation of proton tresinos $p^{*}$'s because there are so many more protons than deuterons in sea water. Now let's consider the situation after all of the deuteron chain reactions have gone to completion, i.e., after many tritium decay times.  In sea water there are roughly 6600 protons for each deuteron and each \mbox{$\alpha$-particle} from the nuclear chain consumes three deuterons (see Figures 5 and 6). Therefore, the proton tresino formation energy per $^{4}$He generated, is about $3\times 6600 \times 3.7\,\,\mathrm{keV}$ or 73 MeV in addition to the 22 MeV from the deuteron chain reactions. This means that some combination of natural U and Th radioactivity (producing only about 5 MeV per $^{4}$He) and the mechanisms just described -- the proton and deuteron tresino formation energy and the deuteron nuclear chain energy together amount to about 95 MeV per $^{4}$He very close to the value estimated by \citet{v&K}.  The two latter components easily account for the observations of the heat and helium discrepancy that has been known now for many years and requiring very little  contribution from U and Th radioactive decay.\,\mbox{Figure 7} is a chart illustrating the various energy source components comparing the observed level, with and without tresino physics (SEEP).
\begin{figure*}[h]
\includegraphics[width=7cm]{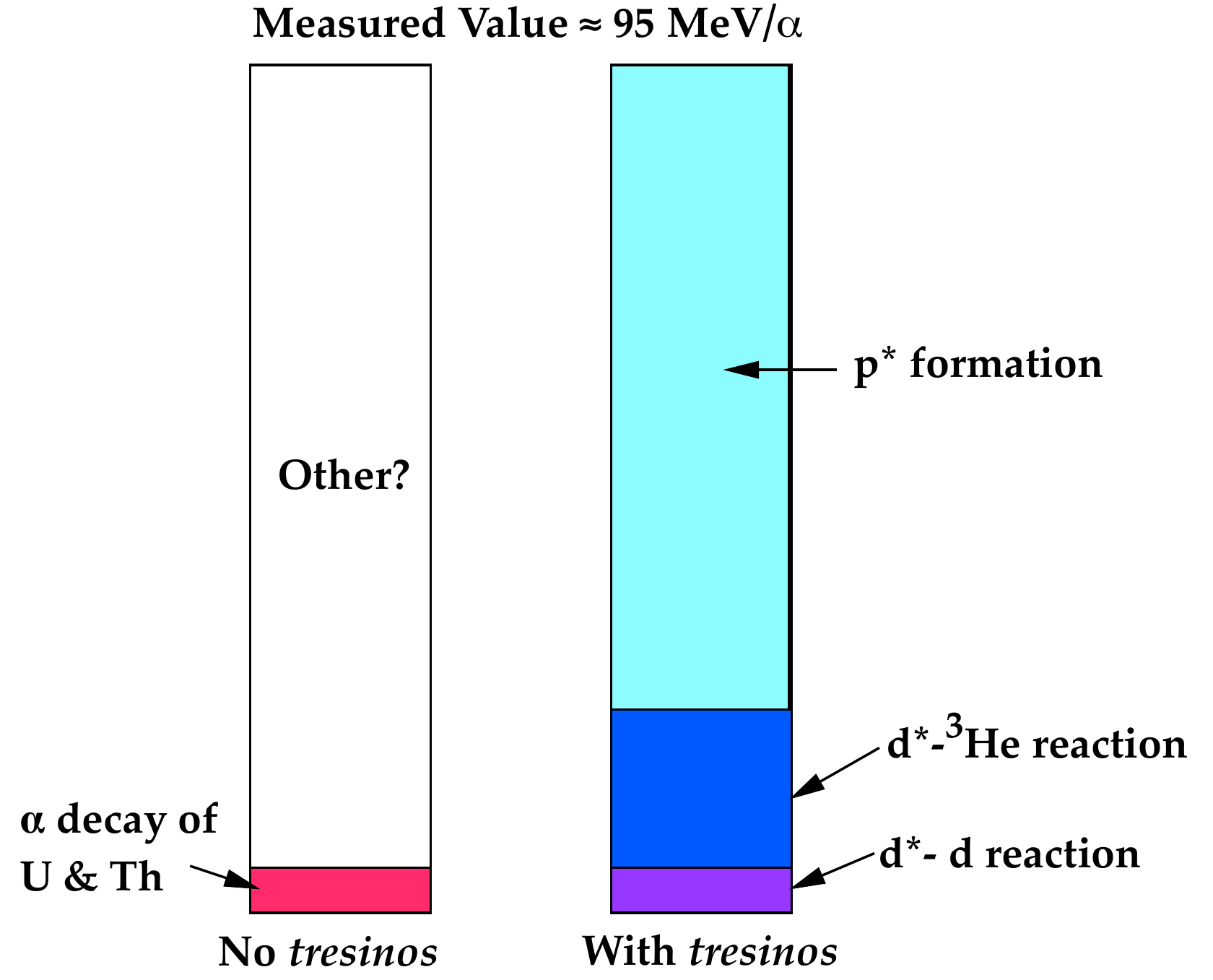}
\caption{A chart comparing the observed heat/helium data with and without tresino physics.}
\end{figure*}
\subsection{Geoneutrinos and the Earth's Heat} 
We note recent measurements of geoneutrinos (\cite{kam} and references therein) have been published that are in direct conflict with our picture. The geoneutrinos putatively the result of U and Th $\alpha$-particle decay sequences followed by numerous $\beta$-decays that release neutrinos. Setting aside issues of statistics from neutrino backgrounds, the authors present numbers of neutrinos detected and numerical models that are consistent with all of the heat, to within a factor of two, having originated from U and Th decays. In model calculations, the authors assumed a bulk Earth (SEEP) model. Their results appear to provide strong evidence in support of the SEEP picture. Indeed, this would certainly be the case if the only nuclear reactions within the Earth came from radioactive decay of U, Th. However, we point out that the deuterium tresino reaction chain presented in this paper has many possibilities for both primary and secondary nuclear reaction channels that produce $\beta$-unstable nuclides and therefore also release neutrinos.  
\par
We have examined many of these reactions and find that the most likely reactions producing neutrinos from $\beta$-unstable nuclides are neutron-transfer reactions of the form $^{\small A}\mathrm{X}(d^{*},p)^{\small A+1}\mathrm{X}$. The latter reactions are similar to those in the deuteron tresino primary reaction chain (\mbox{Figure 5}). For example, consider the reactions on the naturally occuring nuclides $^{27}$Al and $^{23}$Na. These appear to be quite probable because both are plentiful in clays and both produce energetic neutrinos. There are many other ``target nuclides" resident in clays that allow additional neutron-transfer reactions. Furthermore, in addition to the neutron transfer reactions, there are many secondary reactions initiated by the energetic protons and $\alpha$-particles from the primary deuteron reaction chain. Examples of some of these are noted in the following Section. 
\par 
In view of the many deuteron nuclear chain reactions that may yield $\beta$-unstable nuclides (and therefore neutrinos), we think it is erroneous to suggest that the geoneutrino measurements confirm the SEEP picture. 
\section {Discussion and Conclusions}
In our picture, there are three sources of thermal 
energy in the Earth: (i) radioactive decay of U and Th, perhaps 
only a few percent of the total energy, (ii) the deuteron and proton tresino formation energy, the latter being the dominant contribution because of the large amount of hydrogen compared with deuterium in sea water, and (iii) nuclear energy produced in the deuteron-driven chain reactions. 
\par 
This picture of the Earth's heat generation removes the discrepancy between the heat and helium measurements that have been observed for over 30 years.  Furthermore, we find $^{3}$He is a direct consequence of the tresino-driven nuclear chain reactions. Thus, both the remarkably similar $^{3}$He/$^{4}$He ratios at the distant mid-ocean ridges [\cite{Dickin} pg 307] and the higher early-time ratio values, i.e., those close to recent (or imminent) volcanic activity. A recent report by \cite{par} provides clear data showing that monitoring the $^{3}\mathrm{He}/^{4}\mathrm{He}$ ratio is a valuable precursory indication of an imminent volcanic eruption. This is exactly as we would expect within our picture of  energy generation in the Earth. We note again that the relevant time-scale separating these different isotope ratios (close to volcanic activity and far from it) is the tritium decay time. 
\par 
If the tresino hypothesis along with the nuclear chain reactions are the 
 correct physical picture, then there are numerous implications for geophysics 
 research. First of all, most of the heat emanating from the Earth is 
 produced relatively close to the surface and is initiated mostly by entrained surface 
 waters. Since the near-surface heating is not fully spherically 
 symmetric, this could affect the deeper thermal currents which, in 
 turn, might affect the Earth's magnetic field. Asymmetric heating could also be responsible, in part, for some amount of tectonic plate propulsion as well.
\par
An interesting observation relating to volcanic eruptions is the occurance of lightning. Although there have been many suggestions regarding how the lighting builds-up the charge-imbalance, for example [\cite{lightning}], we suggest that residual charge left unbalanced in the tresino formation reactions that drive the eruption could also be involved producing lightning. Of course, experiments would need to examine this possibility.
\par 
Returning to the issue of power levels, we note that the power density in the outer 50 km is quite small (see Appendix B), amounting to only about $10^{-12}$ W/cm$^{3}$ if it were distributed uniformly over the Earth at this depth. It is difficult to imagine how this low power density, if created by radioactive decay of U and Th in the crust, at such low power levels and temperatures, could ever focus this energy into high-power volcanic eruptions. This is to be contrasted with tresino energy generation which is ``naturally focussed" by subduction fissures into which water has seeped or been entrained. So, our picture is that the heat from the Earth is generated primarily under/within the volcanic regions because that is where the water is. And the origin of explosive volcanic events are most likely connected to the rate of inflow or entrainment of water into the accessible deep pressurized fissures. Of course, thermal conduction under the reaction zones will tend to smooth the temperatures away from the zone.
\par 
The Earth's crust and upper mantle are mostly silicate materials, and along with acidified water, they are the major sources of the electron pairs (\em{donors}\rm). If the Earth's water has been consumed over time, then the sea level will have been lowered about 30 meters during the past one billion years if no additional water had accumulated. The effect on ocean level and Earth's hydrogen over the Earth's lifetime will depend in detail on the temperature-dependent rate of tresino formation at higher temperatures.
\par 
Another direct consequence of the deuteron tresino reaction chain results from the flux of energetic protons and \mbox{$\alpha$-particles} localized close to the deuteron chain reactions (see Figure 5).  Numerous secondary nuclear reactions driven by these energetic particles become possible. Some of them may already have been observed in nobel gas isotopic anomalies, but not realized as such. One example is the alpha capture reaction on oxygen $^{16}$O($\alpha,\gamma$)$^{20}$Ne that results in an excess of the isotope $^{20}$Ne compared to $^{22}$Ne. Two other examples, assuming that sulfur and chlorine may be present in the reaction zone, are the reactions $^{36}$S($\alpha,\gamma$)$^{40}$Ar and $^{37}$Cl($\alpha,p$)$^{40}$Ar.  Both of these reactions would create substantial amounts of $^{40}$Ar in excess of $^{36}$Ar. The secondary reactions from the deuteron reaction chain remove the need to invoke atmospheric mixing in the case of Ne and/or the need to invoke ``primordial'' $^{40}$Ar in the case of Ar. Some secondary reactions not only introduce isotope anomalies, some of them are exothermic as well, hence adding to the energy produced in the primary deuteron reaction chain. We note that the ``primordial'' proposal has  been invoked often to account for various anomalies in released gases and recovered nuclides. In particular, $^{3}$He itself has often been proposed to be ``primordial''  [\cite{Dickin} pg 293], or possibly from space [\cite{anderson}]. Some amount of uncertainty may be resolved as a result of the deuteron nuclear chain reactions. 
\par 
We note that \cite{Moreira} present data showing a correleation between ``excess" rare gas (including Ar) isotopic anomalies and $^{3}$He notably one of the nuclides in the primary deuteron nuclear reaction chain. This correlation is exactly what would be expected -  the presence of the $^{3}$He esatblishes that the deuteron nuclear chain is operative and therefore the secondary reactions will generate the isotopic anomalies. The nobel gases observations have provided direct insight into the deuteron nuclear reaction chain because the gases are released but certainly many other isotopically modified nuclides remain in the magma in which they are produced.
\par 
Finally, we note that the deuterium is ``consumed" faster than is the hydrogen (note the lower left panel in Figure 6) in the reaction chains. This disparity implies that the $d/p$ ratio will have been slowly decreasing over the history of the Earth and might impact the difference observed when comparing isotope compositions of terrestrial water and water from comets [\cite{DBock}].
\par
 In this paper, we have tried to show that many otherwise strange, perhaps conflicting, geophysical observations are consistent with heat being produced by proton tresino formation and deuteron tresino-driven primary and secondary nuclear reactions. If this picture is correct then, over time, the nuclear reactions of the tresino chains may have also modified the isotopic and chemical composition of the near-surface Earth. 
\section*{\hspace*{0mm}Appendix  A1 Deuteron Tresino Reaction Rates}
In this Appendix, we derive simplified (approximate) model reaction rates required for later use in reaction rate equations. First consider the \mbox{\em{d-$d^{*}$}\rm} collision in Section 4. The two particles are first accelerated toward each other but at some 
point the electron shielding of the tresino is lost and the two 
particles decelerate. Because of the electron shielding of the tresino, 
the two nuclei can move to within a fraction of a Compton, and at the distance of closest 
approach, a neutron may be transferred accessing one branch of the usual 
\em{d-d}\rm \,\,nuclear reaction. At high collision energies this is a 
straightforward nuclear reaction calculation [\cite{vol}]. However, at the low, almost zero, energies of our collisions, the estimation of the reaction rate is somewhat different. Therefore, we follow a standard derivation 
from \cite{R&R} pg 155-159, but modify it slightly to take account of the difference. The reaction rate may be written as 
\begin{displaymath}
 \hspace{2cm}   r_{dd^{*}}\,=\,n_{d}\,n_{d^{*}}\,\sigma\,V_{s}
    \end{displaymath}
where $\sigma(E) =\, S(E)\,G(E)\,/E $, and 
$V_{s}\,=\,8\times 10^{3}\,T$ $^{1/2}$ ($^{o}\mathrm{K}\rm$) is the speed of sound. Here, $S(E)$ and 
$E$ are the astrophysical $S$-factor and the center of mass energy, respectively. 
The center of mass energy is easily shown to be $3.7\,f^{-1}$\,keV, 
where $f$ the fraction of a Compton at 
which the electron shielding by the tresino is interrupted. 
Usually one would use the Gamow factor $G(E)$ 
and the center of mass energy to complete the estimated reaction 
rate. However, we take \mbox{$S(E)=39$\,keV-barns}, which is
the strictly nuclear part of the neutron transfer branch of the \em{d-d}\rm 
\, reaction \citep[p.~338]{R&R} and replace the Gamow factor times $f$ by a single parameter $P$, a neutron ``tunneling'' parameter. We note that the slow collision between the $d$ and $d^{*}$ will polarize the two deuterons such that their protons remain as far apart as possible. Substituting numerical values, we have  
 \begin{displaymath}
\hspace{2cm}  r_{dd^{*}}\,=9\times 10^{-20}\,P\,\,T^{1/2}n_{d}\,n_{d^{*}}.
    \end{displaymath}  
A similar derivation for the \mbox{ \em$d^{*}$\rm-\em{$^3\mathrm{He}$}\rm}\,  reaction (Section 4, reaction (4)) results in the equation below. To keep the notation less ``cluttered'', we introduce the following changes: $^3\mathrm{He} \equiv 3 $, and    
$^{4}\mathrm{He} \equiv 4$. 
\begin{displaymath}
\hspace{2cm}    r_{{d^{*}3}} \,=1.4 \times 
    10^{-17}\,T^{1/2}\,n_{d^{*}}\,n_{3}\,P. 
\end{displaymath}
\vspace{0.5cm}
\section*{Appendix A2: Infinite-Medium Reaction Rate Equations}
The so-called infinite-medium case in which no particles enter or leave the system (a large system) is the most easily modeled so we focus on this case. Other scenarios, in which some tresinos escape from the system 
after depositing their formation energy, but before they react 
further, can be modeled by similar techniques.
\par 
If there are no particle losses by diffusion, e.g., a sufficiently 
large system, then the chains of self-assembled neutron transfer reactions (Section 4, reactions (2--4)) above can be 
written as a sequence of coupled ordinary differential equations for 
the various species present at any time. We have already given the 
rates, the last two equations in \mbox{Appendix A1}, for tresino formation and the various collisions where $\eta$ is the fraction of ion collisions that produce tresinos, $P$ is the neutron tunneling parameter, and $T(^{o}\mathrm{K})$ is the temperature.
\par 
It is useful, in what follows, to refer to Figure 5, the illustration 
of the proton and deuteron tresino formation and reaction chains. 
Notice the two possible branches of the \em{d-$d^{*}$\rm\, reaction with 
probabilities \,$\epsilon$\, and $(1-\epsilon)$.  We have introduced the $\epsilon$ branching possibility that we adjust in \mbox{Section 5} to be consistent with geophysical observations.
\par
 In terms of these rates, notice that one deuteron is lost 
to $d^{*}$ production and another is lost during a  \em{d-$d^{*}$\rm \, reaction 
resulting in the equation: 
\begin{center}
 $dn_{{d}}/dt\, =\, -\,\,r_{d\,ee}\, - \, r_{d\,d^{*}}$.
\end{center}
Also, \bf{\em{\Large{\color{yellow}{ee}}}}\rm\, donor ions are lost to  $p^{*}$ and $d^{*}$ production as:
\begin{center}
$dn_{\mathrm{ee}}/dt\, =\,-\,r_{d\,ee}\,-r_{p\,ee}$.
\end{center}
For the $d^{*}$ particles, one $d^{*}$ is produced in a collision 
with an $ee$ ion and one is lost to tritium production and another to 
the 3\em{-$d^{*}$}\rm reaction giving: 
\begin{center}
$dn_{d^{*}}/dt\, =\, r_{d\,ee}\,-\, r_{d\,d^{*}}\,-\, r_{3\,d^{*}}$.
\end{center}
For the ``free'' triton production:
\begin{center}
$dn_{t}/dt\,= (1-\epsilon)\, r_{d\,d^{*}}\, - \,n_{t}/\tau$,
\end{center}
where $\tau$ is the tritium decay constant and $\epsilon$ is 
the fraction of \, \em{d-$d^{*}$\rm \, neutron transfer reactions that result in 
a tritium tresino, $t^{*}$.  A triton tresino $t^{*}$ is produced in a fraction, $\epsilon$, of the 
 \em{d-$d^{*}$\rm \, collisions, whereas a fraction, $(1-\epsilon)$, of the  \em{d-$d^{*}$\rm \, collisions produce a ``free'' triton, $t$. For the triton tresino production and loss: 
\begin{center}
$dn_{t^{*}}/dt\,= \,\epsilon \,\,r_{d\,d^{*}}\, -  \, 
    n_{t^{*}}/\tau$.
\end{center}
The $^{3}\mathrm{He}$ particles are produced in beta-decay of the  
tritons and lost to ${^{4}\mathrm{He}}$ production as: 
\begin{center}
$dn_{3}/dt\, =\,n_{t}/\tau\, + \,n_{t^{*}}/\tau\,-\, \,r_{3\,d^{*}}$.
\end{center}
And finally, ${^{4}\mathrm{He}}$, the alpha particles are produced in  
3-$d^{*}$ neutron transfer reactions: 
\begin{center}
$dn_{4}/dt\,= \,r_{3\,d^{*}}$.
\end{center}
The infinite-medium rate equations are gathered together and displayed in \mbox{Section 4}.

\section*{Appendix B:  Simple Earth Heat Flow Calculation}
From both surface and seismic-wave measurements, geophysicists have developed a picture of the interior of the Earth, leading to density, pressure, and temperature profiles.  It has been known for some time that the heat flowing from the Earth comes primarily from the upper mantle and crust, but how much of  the overall production ($\approx 44$ TW) is produced there? 
\begin{table*}[t]
\caption{Table B1}
\begin{tabular}{ccccc}
\hline
Zone depth(km)    &  \hspace{0cm}   $\kappa$(W/$^{o}$K-km)    & \hspace{0cm} $H$(W/km$^{3}$)    & 
   \hspace{0cm} $T_{f}$($^{o}$K)    &  \hspace{0cm}   $dT_{f}/dz$($^{o}$K/km)   \\ 0-50 &\hspace{0cm} 4000 &\hspace{0cm} 1000  & \hspace{0cm}990 & \hspace{0cm}7.5\\
50-200 &\hspace{0cm} 4500 & \hspace{0cm}195 & \hspace{0cm}1630 & \hspace{0cm}1.0\\
\hline
\end{tabular}
\end{table*}

\par
 The temperature profile of the Earth's interior is determined by heat production and heat transfer processes, which can be analyzed by an appropriate energy-conservation equation.   In the near-surface region under steady-state conditions, this is just the familiar thermal conduction equation \mbox{$\kappa \, d^{2}T/dz^{2} = - H$} where $T(z)$ is the temperature, $\kappa$ is the thermal conductivity  and $H$ is the power 
produced per unit volume. Using the temperature profile for the first 200 km depth in the Earth from \citet{K&O} or \citet{Stacey}, the thermal conductivities from Stacey, and $T(z)$ and $dT/dz$ surface values of 300\,{$^{o}$K} 
and 20 {$^{o}$K}/km. It is straightforward to calculate the data presented in Table B1.  Note that $f$ refers to the value at the end of the zone. With the values of $H$ listed in Table B1, we find the $T_{f}$ and $dT_{f}/dz$ values shown and they agree quite well with those of the temperature profile given by Stacey. Also, with these values for $H$, the heat flow from the two zones are 25 TW and 14 TW, respectively. The total for these two zones is almost all of the observed 44 TW. However, note that this simple heat flow exercise is independent of the source, or sources, producing the energy. 


\end{document}